\documentclass[prb,aps,twocolumn,superscriptaddress,showpacs,floatfix,letterpaper]{revtex4-1}  
\usepackage{hyperref}
\usepackage{amssymb}
\usepackage{amsmath}
\usepackage{bbm}
\usepackage{graphicx}
\usepackage{epsfig}
\usepackage{epstopdf}
\usepackage[usenames]{color}
\usepackage{soul}
\usepackage{lipsum}
\usepackage{bbold}
\usepackage{afterpage}
\usepackage{cancel}
\hypersetup{
  colorlinks = true, 
  urlcolor   = blue, 
  linkcolor  = blue, 
  citecolor  = blue  
}

\begin{document}
\bibliographystyle{apsrev4-1}

\title{Lattice Energetics and  Correlation-Driven Metal-Insulator Transitions:  The  Case of Ca$_2$RuO$_4$}
\author{Qiang Han}
\affiliation{Department of Physics, Columbia University, New York, New York 10027, USA} 
\author{Andrew Millis}
\affiliation{Department of Physics, Columbia University, New York, New York 10027, USA} 
\affiliation{The Center for Computational Quantum Physics, The Flatiron Institute, New York, New York 10010, USA}

\date{\today}

\begin{abstract}
This Letter uses density functional, dynamical mean field, and Landau-theory methods to elucidate the interplay of electronic and structural energetics in the Mott metal-insulator transition.  A Landau-theory free energy is presented that  incorporates the electronic energetics, the coupling of the electronic state to local distortions and the coupling of local distortions to long-wavelength strains. The theory is applied to Ca$_2$RuO$_4$. The change in lattice energy across the metal-insulator transition is comparable to the change in electronic energy. Important consequences are a strongly first order  transition, a sensitive dependence of the phase boundary  on pressure and that the geometrical constraints on in-plane lattice parameter associated with epitaxial growth on a substrate typically change the lattice energetics enough to eliminate the metal-insulator transition entirely. 
\end{abstract}
\pacs{71.27.+a,75.50.Cc,72.15.Eb}
\maketitle
Many materials exhibit "Mott" metal-insulator transitions, primarily driven by electron-electron interactions \cite{Imada98} but also involving changes in atomic positions. In  the rare earth titanates  and vanadates, the distortion associated with  the insulating phase is  a GdFeO$_3$-type octahedral rotation\cite{PhysRevB.68.060401,Pavarini04}, in the rare earth manganites, it is an approximately volume-preserving even-parity octahedral distortion\cite{Millis95,Gosnet03,Tokura06}: in the perovskite  nickelates,  a two sublattice disproportionation of the mean Ni-O bond length \cite{Torrance92,Alonso99,Fernandez-diaz00,Medarde09} and in VO$_2$  a V-V dimerization \cite{Morin59}. In other materials including Ca$_2$RuO$_4$ \cite{Maeno97} and V$_2$O$_3$ \cite{McWhan69} the metal-insulator transition occurs simultaneously with a crystal symmetry-preserving change of atomic positions.  The association of metal-insulator and structural transitions suggests the possibility of tuning electronic behavior by strain \cite{VO2}, epitaxial growth, or  "nonlinear phononic" effects arising from  intense terahertz radiation\cite{Cavalleri1,Cavalleriprl,hiddenphase}.    

While  electronic aspects of the Mott transition are becoming well understood, and energies, forces, and many-body structural relaxation  are now  available in beyond  density functional  frameworks such as the density functional plus dynamical mean field methodology  \cite{Park14,Leonov14,Haule16}, the interplay between the lattice and electronic energetics has yet to be fully unraveled.  A physical basis for interpreting the calculations and the experiments remains to be defined and the magnitude of the lattice contribution to the energetics of the transition has yet to be determined. Here we argue that the  key point is that the electronic transition couples directly to  local atomic configurations such as octahedral rotations and transition metal-oxygen bond lengths, which in turn  couple directly to externally controllable variables such as strain and pressure. The response of the material to these stresses defines a lattice stabilization energy, which can in fact be large enough to dominate the energetics of the transition. 

To quantify these effects  we write an electronic free energy $F^a(\delta \vec{Q})$ that depends on a state variable $a$ labeling whether the material is in the metallic or insulating phase, and on atomic coordinates, labeled by a vector $\delta \vec{Q}$ expressing deviations of atomic positions from a reference configuration.  Expanding in $\delta \vec{Q}$ we obtain

\begin{equation}
F^a(\delta \vec{Q})=F_0^a+\vec{\mathcal{F}}^a\cdot\delta\vec{Q}+\frac{1}{2}\delta \vec{Q}^T\cdot\mathbf{K}^a\cdot\delta\vec{Q}+...
\label{eq:F}
\end{equation}
The electronic state-dependent linear term $\vec{\mathcal{F}}^a$  specifies the force exerted by the electronic state on the atomic degrees of freedom. Typically $\vec{\mathcal{F}}$ couples only to a subset of the lattice degrees of freedom, but this subset is coupled to other lattice coordinates by the quadratic restoring term $\mathbf{K}$.  The ellipsis represents anharmonic terms that are not needed for the considerations of this Letter but may be important in other circumstances \cite{Kanamori65,Millis95}. 

Minimizing the terms  written in Eq. (1) gives $F=F_0^a-\frac{1}{2}\vec{\mathcal{F}^a}^T\cdot\mathbf{K}^{-1}\cdot\vec{\mathcal{F}}^a$ defining the stabilization energy
\begin{equation}
F_{stabil}^a=-\frac{1}{2}\vec{\mathcal{F}^a}^T\cdot\mathbf{K}^{-1}\cdot\vec{\mathcal{F}}^a
\label{eq:Estabil}
\end{equation} 
so that if the lattice is free to relax, the transition between phases $a=1,2$ will occur when $F_0^1+F^1_{stabil}=F_0^2+F^2_{stabil}$ corresponding to a shift in the transition point relative to a frozen lattice calculation and  a lattice change across the transition $\delta \vec{Q}^1-\delta\vec{Q}^2=-\mathbf{K}^{-1}\left(\vec{\mathcal{F}}^1-\vec{\mathcal{F}}^2\right)$. 
\begin{figure}[t]
\includegraphics[width=0.3\columnwidth]{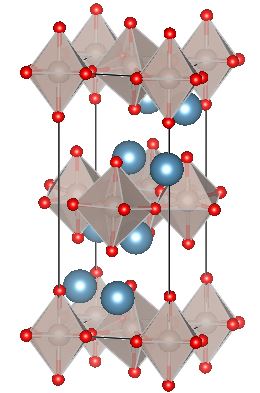}
\includegraphics[width=0.65\columnwidth]{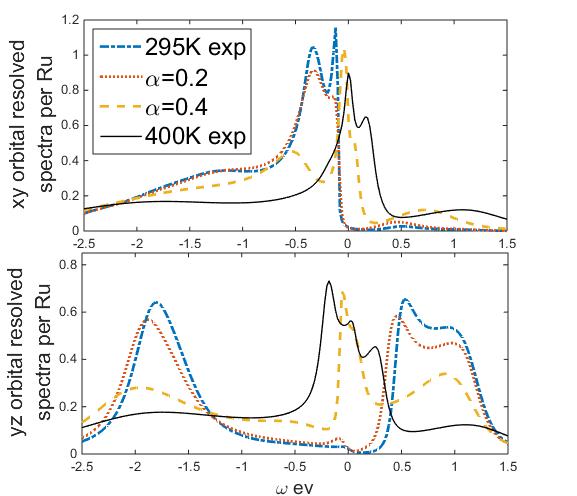}
\caption{Left panel: Representation of the unit cell of Ca$_2$RuO$_4$. Gray balls are ruthenium atoms, red balls oxygen atoms and blue balls calcium atoms. Right panel: Orbitally resolved many-body densities of states for structures interpolating between experimental 295 and 400 K structures. Upper panel: $xy$ orbital; lower panel: $yz$ orbital ($xz$ is very similar). $\alpha=0$ is the 295 K structure; $\alpha=1$ is the 400 K structure; $\alpha=0.4$ is in the metallic phase but very close to the transition point; the $\alpha=0.6,0.8$ spectra are very similar to the 400 K spectra and are omitted for clarity.}
\label{fig:structspect}
\end{figure}

We now apply these generic considerations to  Ca$_2$RuO$_4$, which exhibits a correlation-driven paramagnetic metal to paramagnetic insulator transition as the temperature is decreased below a critical value about 350 K\cite{Maeno97}. The transition is accompanied by a  large amplitude, symmetry preserving lattice distortion  \cite{Braden98,Friedt01}. Below about 140 K there is an onset of antiferromagnetic order\cite{Braden98,Friedt01}, which is not relevant to our present considerations.  Ca$_2$RuO$_4$  crystallizes in a $Pbca$-symmetry structure with four formula units in each crystallographic unit cell. The basic structural unit is the Ru-O$_6$ octahedron; these  form corner-shared planes separated from adjacent Ru-O$_6$ planes by layers  involving Ca atoms. The left panel of  Fig.~\ref{fig:structspect} shows one unit cell with four formula units.  The $Pbca$ structure is derived from the ideal tetragonal $n=1$ Ruddlesden-Popper structure by  rotations of the Ru-O$_6$ octahedrons about the apical Ru-O(2) bonds, tilts of this axis with respect to the Ru-O(1) plane, as well as an additional distortion that makes the two in-plane Ru-O bond lengths slightly different. The apical [Ru-O(2)] and  the average over the two in-plane directions [Ru-O(1)] Ru-O bond lengths are the crucial variables in the electronic energetics. Their values  across the metal-insulator transition are presented in Table~\ref{table:onsiteenergy}.   The bond lengths  continue to evolve as temperature is further lowered through  the insulating phase\cite{Braden98,Friedt01}.  The corner-shared structure implies that if the rotation angles remain fixed, the  Ru-O(1) bond lengths predict the average in-plane lattice parameters.  Density functional calculations show that changes in the rotation angles are negligible for reasonable strains\cite{Supplemental}, so the in-plane Ru-Ru and Ru-O(1) distances are not independent variables.  On the other hand, the $c$-axis stacking of the Ruddlesden-Popper structure means that at fixed $c$-axis lattice constant,  changes in the Ru-O(2) bond length can  be accommodated by a buckling of the Ca-O planes.

We will be interested here in structures where the $c$-axis lattice parameter is relaxed for given values of the octahedral bond lengths.   Thus the lattice degrees of freedom in our theory are the average Ru-O(1) and  Ru-O(2) lengths.  We parametrize the Ru-O bond lengths in terms of changes $\delta x,\delta y,\delta z$ with respect to a reference state, which we take to be the 400 K metallic state, and we express these in terms of the octahedral coordinates
\begin{equation}
\delta Q_0= \frac{1}{\sqrt{3}}(\delta z+\delta x+\delta y)\hspace{0.2in}\delta Q_3= \frac{1}{\sqrt{6}}(2\delta z-\delta x-\delta y)
\label{eq:latticedistortion}
\end{equation}  
which we assemble into the vector $\delta\vec{Q}=\left(\delta Q_3,\delta Q_0\right)$. $\mathbf{K}$ in Eq. (1) is defined from the dependence of energies on $\delta Q_3$ and $\delta Q_0$, with the c-axis lattice constant relaxed for each value of $\delta \vec{Q}$. We used density functional plus U (DFT+U) calculations and observed phonon frequencies (which give energetics of Ru-O bond length changes without lattice relaxation) to estimate the entries of $\mathbf{K}$ (see Supplemental Material\cite{Supplemental}), finding  $K_{33}=17.7$, $K_{03}=7.6$, $K_{00}=46.2$ eV$/\AA^2$ per formula unit. The  observation\cite{PhysRevB.71.245121,PhysRevLett.91.056403,Souliou17} that the changes in optical phonon frequencies across the transition are about $2\%$, justifies the harmonic approximation and the independence of $\mathbf{K}$ on the electronic phase. 

\begin{table}[htbp]
\centering
\caption{Experimentally determined apical [Ru-O(2)] and average in-plane [Ru-O(1)] bond lengths and octahedral distortions [Eq.~(3)] in \AA ~at T=295 K~[\onlinecite{Braden98}] and 400 K~[\onlinecite{Friedt01}],  and occupancy (per spin per atom) of $xy$ ($n_{xy}$), and average of $yz$, $zx$ ($\bar{n}_{yz/zx}$) orbitals  from DMFT calculation  using the experimentally determined lattice structures at each temperature. }
\label{table:onsiteenergy}
\begin{tabular}{c c c c c c c }
\hline
          & RuO(2)    & RuO(1) &$\delta Q_0$& $\delta Q_3$& $n_{xy}$  & $\bar{n}_{yz/zx}$ \\
\hline
400~K & 2.042  & 1.95 & 0.0&0.0& 0.671 & 0.665\\
295~K & 1.995 & 1.99 &0.0196& -0.069&  0.982 & 0.508\\ \hline
\end{tabular}
\end{table}

We now turn to the electronic degrees of freedom. The relevant frontier electronic states are $t_{2g}$-derived  Ru-$4d$ oxygen $2p$ antibonding states which we refer to as Ru $d$ states, following standard practice \cite{Allen96,Ahn99,gorelov2010nature,PhysRevB.91.195149,PhysRevB.93.155103}.  The $t_{2g}$-derived bands are  well separated from the other bands, so we may  focus our treatment of the correlation problem on them, treating the other bands as inert\cite{gorelov2010nature,PhysRevB.91.195149,PhysRevB.93.155103}. The tetragonal symmetry  splits the $t_{2g}$-derived triplet  into a singlet ($d_{xy}$) and a doublet ($d_{xz}$ and $d_{yz}$). The octahedral rotations and other distortions  (angles $\sim 10^\circ$) provide small additional rearrangements of the level structure (in particular lifting the $xz$/$yz$ degeneracy),  but  as long as  the orbitals are defined  with respect to the local octahedral symmetry axes, the deviations from the perfectly tetragonal structure do not significantly affect the on-site level splitting, basic energetics, or assignment of orbital character.  Spin-orbit coupling ($\lambda_{SOC}\approx 0.1$~eV) is important for lower $T$ magnetic properties of the insulating state  \cite{Souliou17,Sutter17,Kim17} but is not relevant to the physics we consider here since the spin-orbit  energy scales are small compared to the orbital level splitting and electron interactions.

We have performed DFT and DFT+DMFT calculations (see Supplemental Material\cite{Supplemental2}). We find, in agreement with previous work \cite{gorelov2010nature}, that a calculation at room temperature with the experimentally determined 400~K structure produces a moderately correlated metallic solution while using the 295~K structure produces a Mott insulator. The metallic state is characterized by an approximately equal occupancy of the three $t_{2g}$ orbitals. The approximately equal orbital occupancy is not protected by any symmetry and is  due to the strong electron-electron scattering.  The insulating state is orbitally disproportionated,  with an essentially fully filled $xy$ band and half filled, much narrower, $xz/yz$ bands with  upper and lower Hubbard bands separated by a gap (Fig.~\ref{fig:structspect} right panel with blue dashed lines).  Calculated orbital occupancies are given in Table ~\ref{table:onsiteenergy}.

\begin{figure}[t]
\includegraphics[width=1.0\columnwidth]{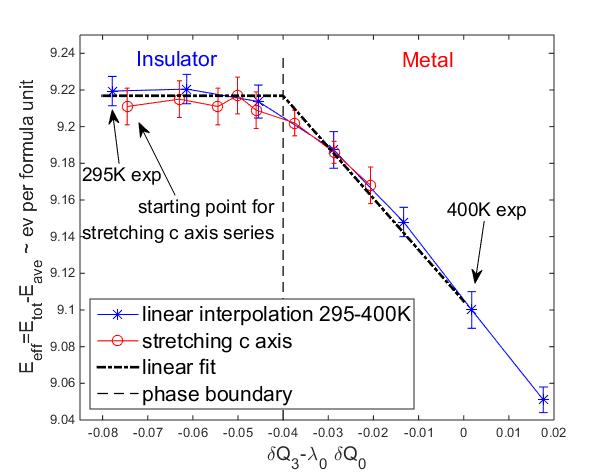}\\
\caption{Electronic energy of correlated bands  $E_{eff}=E_{corr}-\epsilon_{ave}N_{tot}$ plotted against a linear combination of octahedral parameters with $\lambda_0=0.45$ and calculated using  DFT+DMFT for two series of structures: the linearly interpolated structures between the experimentally observed metallic 400 K and insulating 295 K structures (solid points, blue) and a series obtained by starting from an relaxed insulating structure with a=b=5.44~\AA~ and stretching the $c$-axis (open symbols, red). The bold dashed black line stands for the linear fit in Eq.~(4) and the light dashed line shows the phase boundary. The error bars are statistical errors from the Monte Carlo solution of the DMFT equations.}
\label{fig:Evslattice}
\end{figure}

The right-hand panels of Fig.~\ref{fig:structspect} present the orbitally resolved densities of states obtained from DFT+DMFT calculations at room temperature, performed for a series of structures linearly interpolated between the $T$=295 ($\alpha=0$) and $T$=400 K ($\alpha=1$)  structures. As the interpolation parameter $\alpha$ changes from $1$ to $0.4$, the state remains metallic but the  bands and occupancies ($n_{xy},n_{xz},n_{yz}$) change from $\approx(4/3,4/3,4/3)$ to $\approx(5/3,7/6,7/6)$. A first order MIT occurs as $\alpha$ is decreased below a critical value  $\approx 0.4$. Further changes of structure within the insulating phase ($\alpha= 0,~ 0.2$) do not affect the orbital occupancies but do lead to an approximately $0.1$~eV shift upward of the $xz/yz$ band relative to the $xy$ band. We have also performed calculations in which one starts from the DFT+U relaxed insulating phase atomic positions  with in-plane lattice constants fixed to 5.44~\AA~ and the $c$-axis parameter is then  gradually stretched. The results are very similar to the first group. Although the transition is first order we have not observed  coexistence  of metal and insulator phases  at any of the lattice configurations we have studied.  

\begin{figure*}[!htb]
	\includegraphics[width=0.95\textwidth]{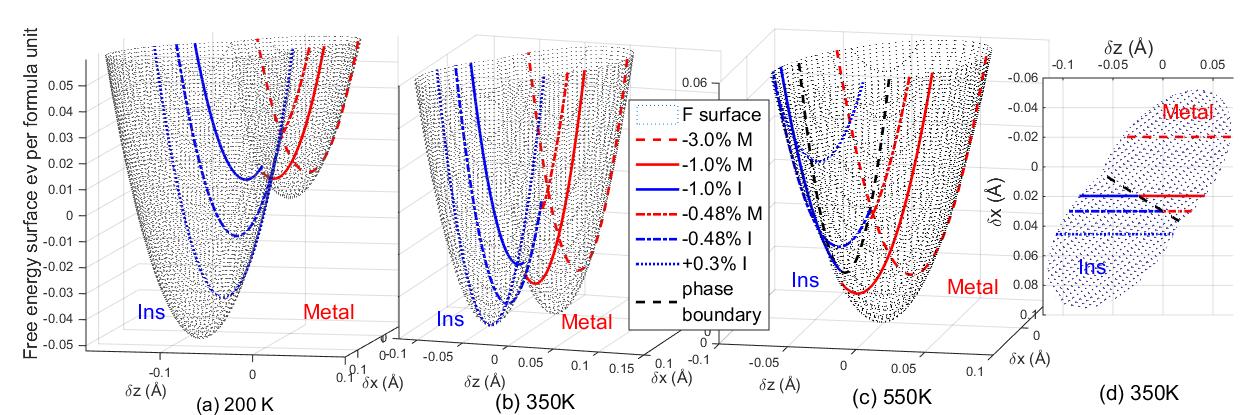}
	\caption{Free energy surfaces computed for unconstrained bulk Ca$_2$RuO$_4$ at temperatures 200 (a), 350 (b), and 550~K(c), along with projection of the 350 K surface onto the x-z plane (d). The black dashed lines in panels (c) and (d) show the metal-insulator phase boundary. The solid and dashed lines show the structural trajectories which the system can explore for films grown under the epitaxial  strain conditions given in the legends. The blue and red regions of the lines indicate insulating and metallic regions respectively.  }
	\label{fig:Fsurface1}
\end{figure*}

Fig.~\ref{fig:Evslattice} plots the DFT+DMFT energy of the correlated bands (obtained as described in the Supplemental Material\cite{Supplemental2} for interpolated and $c$-axis stretched structures) against a linear combination of octahedral parameters [Eq.~(3)].  
\begin{equation}
\footnotesize
\begin{split} 
E_{eff}=&E_{corr}-\epsilon_{ave}N_{tot}\\
=&E_0-\mathbf{\mathcal{F}}_3(\delta Q_3-\lambda_0\delta Q_0-\delta Q_c)\Theta(\delta Q_3-\lambda_0\delta Q_0-\delta Q_c)
\label{eq:Eeff} 
\end{split} 
\end{equation}
Here $\epsilon_{ave}$  is the orbitally averaged on-site energy from  MLWF fits to the converged DFT band structures; $N_{tot}=4$, and $\epsilon_{ave}N_{tot}$ basically represents the insulating phase electron energy up to a constant. The particular linear combination with $\lambda_0=0.45$ is chosen so that the data from the two different families of  structures (which change the bandwidth and octahedral distortion in different proportions) collapses in both the insulating and metallic phases. The dependence on $\delta Q_3$ reflects the relation between the octahedral shape and the orbital splitting. The dependence on $\delta Q_0$ reflects the change in bandwidth. We emphasize that the insulating (metallic) state is only stable for $\delta Q_3-0.45\delta Q_0<(>)\delta Q_c\approx -0.04~\AA$ (we expect $\delta Q_c$ depends on $U,~J$). Apart from some rounding in the immediate vicinity of the transition, the energy is a linear function of the relevant combination of the structural parameters, with a difference in slope between phases. The curvatures $\partial^2 E_{eff}/\partial \delta Q^2$ in two phases are difficult to determine accurately from these calculations but are small enough compared with the $\mathbf{K}$ that any change in the ${\mathbf K}$ across the phase boundary is negligible (details are in the Supplemental Material\cite{Supplemental2}).

The choice of variables in Fig.~\ref{fig:Evslattice} fixes the change in force across the transition as $\mathcal{F}_3=\mathcal{F}^I_3-\mathcal{F}^M_3=2.8$ and $\mathcal{F}_0=\mathcal{F}^I_0-\mathcal{F}^M_0=-0.45(\mathcal{F}^I_3-\mathcal{F}^M_3)=-1.3$~eV/\AA.  Within the assumptions made here, the dependence of the  insulating phase energy on $\delta\vec{Q}$  is essentially independent of temperature. However,  as temperature is further lowered through the paramagnetic insulating phase to the  AFM phase transition, an approximately linear evolution of the Ru-O bonds lengths is observed\cite{Braden98,Friedt01}, indicating an approximately $T$-linear dependence of the insulating-state force. Linearly extrapolating the Ru-O bond lengths measured in experiments \cite{Braden98,Friedt01} at 180, 295 and 350~K to 0~K yields results within 14\% of our calculated values. We therefore believe that the single-site DMFT theory used here is a good representation of $T\rightarrow 0$~K energetics and that the temperature dependence is due to  entropic terms arising from a combination of intersite effects missing in the single-site approximation used here,  spin orbit effects which change the on-site multiplet structure and lattice contributions. We model these effect by a phenomenological linear term in $\vec{F}$, so 
\begin{equation}
\vec{\mathcal{F}}=\left(\begin{array}{c}\mathcal{F}_3 \\\mathcal{F}_0\end{array}\right)\left(1-0.0017T[K]\right)
\label{eq:Fexp}
\end{equation}

The consistency of the model can be verified via a computation of the pressure dependence of the transition. This is obtained by adding to Eq.~(1) a term $+PdV=P\frac{1}{4}(ab\delta c+ac\delta b +bc\delta a)=P(\beta_3 \delta Q_3 +\beta_0 \delta Q_0)$ with $(\beta_3,\beta_0)=(-0.3281, -0.1861)$~eV/(GPa$\cdot$\AA~ formula~unit), so that applied pressure is in effect a linear term shifting the position and value of the energy minimum. We find $P_c=3.6-0.011T~(Gpa)$ which is comparable to $P_c^{exp}\approx 2.3-0.006T(Gpa)$ fitted from published data\cite{PressureinducedMIT}. More details are in the Supplemental Material\cite{Supplemental2}.

In Fig.~\ref{fig:Fsurface1} we plot the free energy landscape at different temperatures in the plane of Ru-O bond length coordinates $\delta x=\frac{1}{\sqrt{3}}\delta Q_0-\frac{1}{\sqrt{6}}\delta Q_3$ and $\delta z=\frac{1}{\sqrt{3}}\delta Q_0+\frac{\sqrt{6}}{3}\delta Q_3$, using force terms estimated in Eq.~(5). We chose the metallic state at $T>T_{M-I}$  as the reference. At high temperature, there is no global minimum in the insulating phase. For $T\leq T_{M-I}$, an insulating energy minimum as in Eq.~(2) appears and becomes more stable. The stabilization energy defined in Eq.~(2) is $\approx -0.048$~eV/Ru at $T_{M-I}$.

We now turn to epitaxially grown films. While epitaxial films are strained with respect to bulk, strain is not the key issue. Rather, the tight association of the in-plane lattice parameter and the Ru-O(1) bond length means that epitaxy implies a constraint: instead of freely minimizing Eq.~(1) over the full space of structural variables, the system can explore only a one-dimensional cut across the energy landscape, corresponding to a fixed Ru-O(1) bond length. The solid and dashed lines in Fig.~\ref{fig:Fsurface1} show the one-dimensional cuts which can be explored under different epitaxy conditions.  Because the curves typically do not pass near the global minimum,  the phase transition becomes much more expensive and in most cases is eliminated.  Only in a small range of compressive strains around $-1.0\%$ (relative to 295~K structure)  can a metal-insulator transition occur in a reasonable temperature range.   For a larger compressive strain the system is always a metal while for a tensile or small compressive strain the material is always an insulator. This  is consistent with recent experimental observations \cite{Nair,Straineffectexp} that thin films of Ca$_2$RuO$_4$  grown epitaxially on NdGaO$_3$ (+0.3$\%$ strain) and NSAT (-0.48$\%$) remain insulating up to 550~K while films grown on NdAlO$_3$ (-3.0$\%$) remain metallic down to lowest temperature. Only films grown on LaAlO$_3$ (-1.6\%)  exhibit a transition to a weakly insulating phase at  $T\approx 200$~K.

In summary, we demonstrated the importance of lattice energetics in the Mott metal-insulator transition, elucidating the crucial interplay between the local octahedral distortions and long wavelength strains, and the previously unappreciated role of epitaxial constraints.  We focused on Ca$_2$RuO$_4$, which has two simplifying features: the metal and insulator have the same symmetry and octahedral rotations are of minor importance, so the order parameter couples linearly to strains and the in-plane Ru-O bond lengths determine the Ru-Ru spacing. Performing a complete DFT+DMFT structural relaxation study and providing a less phenomenological treatment of the electronic and, especially, lattice entropies  are also  important directions for future research.  Most importantly, a generalization of the theory to cases where octahedral rotation is important (perovskite titanates and vanadates) or the insulating phase breaks a translation symmetry (manganites and nickelates) so that strain couples via nonlinear terms in the elastic theory, is urgently needed.

We thank H. Nair, D.Schlom, Jacob Ruf and K.Shen for sharing  data in advance of publication and for helpful discussions. The DMFT calculations used codes written by H. T. Dang and were performed on the Yeti HPC cluster at Columbia University. This research is supported by the Basic Energy Science Program of the Department of Energy under Grant No. ER-04169 and Cornell Center for Materials Research with funding from the NSF MRSEC program (DMR-1120296).

\bibliography{bibofCa2RuO4_0714}
\end{document}